\documentclass{osa-article}

\journal{oe}
\articletype{Research Article}
\usepackage{graphicx}
\usepackage{url}
\usepackage{appendix}

\usepackage[euler]{textgreek}

\graphicspath{{Figures/}}

\begin{document}

%\preprint{}

\title{Optimization of photoluminescence from W centers in silicon-on-insulator} %Title of paper

\author{Sonia M. Buckley,\authormark{1,*} Alexander N. Tait,\authormark{1} Galan Moody\authormark{1,3}, Bryce Primavera,\authormark{1}, Stephen Olson\authormark{2}, Joshua Herman\authormark{2}, Kevin L. Silverman\authormark{1}, Satyavolu Papa Rao\authormark{2},
Sae Woo Nam\authormark{1}, Richard P. Mirin\authormark{1}, Jeffrey M. Shainline\authormark{1}}

\address{\authormark{1}National Institute of Standards and Technology, 325 Broadway, Boulder CO 80305, USA\\
\authormark{2}SUNY Polytechnic Institute, 257 Fuller Road, Albany, NY 12203, USA\\
\authormark{3}Currently at the Department of Electrical and Computer Engineering, University of California Santa Barbara, Santa Barbara, CA 93106}

\begin{abstract}
W centers are trigonal defects generated by self-ion implantation in silicon that exhibit photoluminescence at 1.218 \textmu m. We have shown previously that they can be used in waveguide-integrated all-silicon light-emitting diodes (LEDs). Here we optimize the implant energy, fluence and anneal conditions to maximize the photoluminescence intensity for W centers implanted in silicon-on-insulator, a substrate suitable for waveguide-integrated devices. After optimization, we observe near two orders of magnitude improvement in photoluminescence intensity relative to the conditions with the stopping range of the implanted ions at the center of the silicon device layer. The previously demonstrated waveguide-integrated LED used implant conditions with the stopping range at the center of this layer. We further show that such light sources can be manufactured at the 300-mm scale by demonstrating photoluminescence of similar intensity from 300 mm silicon-on-insulator wafers. The luminescence uniformity across the entire wafer is within the measurement error.
\end{abstract}

\section{Introduction}
\label{sec.introduction}

The indirect bandgap of silicon leads to inefficient optical transitions at room temperature, and therefore light sources on a silicon platform typically involve heterogeneous integration of other materials, such as compound semiconductors. A low-cost silicon-based light source would be of enormous benefit for optical interconnects and communications, but despite much effort \cite{Zhou2015,Liang2010}, it remains elusive. The diverse list of applications for silicon photonics includes those utilizing photonic circuits with superconducting electronics or detectors, for which cryogenic operation is required. In particular, superconducting optoelectronic hardware has been proposed as a platform for neuromorphic supercomputing \cite{Shainline2017a,shbu2018} as well as integrated-photonic quantum computing \cite{Rudolph2016}. For these cryogenic applications, light emission processes that only exist (or have much higher efficiency) at cryogenic temperatures are useful. This area still has room for exploration, as far less attention has been given to the development of silicon-based light sources that are intended to operate at low temperature.  These applications also benefit from leveraging processes that are compatible with standard semiconductor microfabrication for reliability and scaling. 

Point defects in indirect bandgap semiconductors can act as radiative recombination centers. Light-emission processes based on point defects often show significantly higher efficiency at low temperatures\cite{Davies1989,Recht2009}. A large number of emissive defects in silicon exist and have been studied \cite{Davies1989, Shainline2007}. Electrically-injected light-emitting diodes (LEDs) based on implanted defects and dislocations in silicon have been demonstrated \cite{Bradfield1989,Libertino2000,Cloutier2005,Bao2007,Lourenco2008,LoSavio2011}. However, most of these studies were focused on achieving room temperature operation. In this study, we optimize performance at cryogenic temperatures, where the LEDs are intended to operate for compatibility with superconducting technologies. Unlike light sources fabricated on more exotic substrates, light sources based on defects in silicon can be fabricated easily in standard silicon wafers, with both silicon and silicon-on-insulator (SOI) wafers available up to 300 mm.

The W center \cite{Davies1987} is an emissive center generated by self-ion implantation of silicon ions \cite{Yang2010}. It is believed to have a trigonal geometry \cite{Davies1987} and be made up of interstitial silicon atoms \cite{Carvalho2005a,Santos2016,Aboy2018}. We chose to study the W center over any of the large number of other emissive centers in silicon for several reasons. It is relatively well studied and easy to make, requiring a single implant and anneal step. Electrical injection of W-center LEDs has been demonstrated. LEDs based on ensembles of W centers implanted in the $i$-region of a $p$-$i$-$n$ diode have been demonstrated \cite{Bao2007}, and we have previously demonstrated waveguide-coupled W-center LEDs in a cryogenic optical link with superconducting single-photon detectors \cite{Buckley2017}. The W center emits at a convenient wavelength for silicon photonics. The zero phonon line for the W center is centered at 1.218 \textmu m, which falls below the silicon indirect bandgap while being above the SiO$_2$ phonon band. 

Despite previous studies in a SOI substrate \cite{Wang2011,Buckley2017}, an in-depth study of the optimal conditions for maximizing light emission from ensembles of W centers in SOI has not been performed. SOI is the workhorse for silicon photonics, as the mode confinement provided by the buried oxide layer is necessary for waveguides. Previous waveguide integrated LEDs exhibited an external efficiency\cite{Buckley2017} of 5$\times10^{-7}$, while similar measurements on W-center LEDs in bulk silicon yielded external efficiency\cite{Bao2007} of 10$^{-6}$. The efficiency value in Ref.\,\cite{Buckley2017} includes losses due to the poor waveguide coupling, the device resistance, and the low quantum efficiency of the waveguide-coupled diode itself.  Here, we show that the photoluminescence (PL) intensity can be improved by over two orders of magnitude through optimization of the implant conditions. In particular, we have studied the effect of implantation energy, fluence, and annealing conditions for optimization of PL from SOI wafers. We expect that this will translate into a similar increase in efficiency in electrically injected LEDs. This study may also provide helpful information to those hoping to further understand the properties of W centers in silicon. 

\section{Characterization and setup}
\label{sec.characterization}

Implants were initially performed using a commercial ion implantation service in three different wafer types: (A) 1 $\Omega\cdot$cm to 10 $\Omega\cdot$cm 76.2 mm $p$-type (boron doped) silicon, (B) $>$ 10000 $\Omega\cdot$cm 76.2 mm undoped silicon, (C) $p$-type (boron doped) SOI 76.2 mm with a 220 nm silicon layer on 3 \textmu m buried oxide. Wafers were cleaned in a sulfuric acid solution followed by hydrofluoric acid and a de-ionized water rinse. A 7 nm thermal oxide was subsequently grown on the wafers before implant. All implants were performed at room temperature at 7$^\circ$ off normal to prevent channeling. Samples were subsequently annealed in N$_2$ ambient at 250$^\circ$C for 30 minutes unless otherwise stated.  Initial characterization of the W centers was performed in a continuous-flow cryostat with a minimum temperature of 4.2 K. The samples were pumped using a continuous-wave (CW) HeNe laser at 632 nm through a 0.6 NA 20$\times$ objective lens with a 20 mm working distance, unless otherwise noted. The W center PL was collected through the same lens. The PL was then passed through a grating monochrometer and detected on a liquid-nitrogen-cooled linear InGaAs photodiode array.

\begin{figure}
\centering\includegraphics[width=11.4cm]{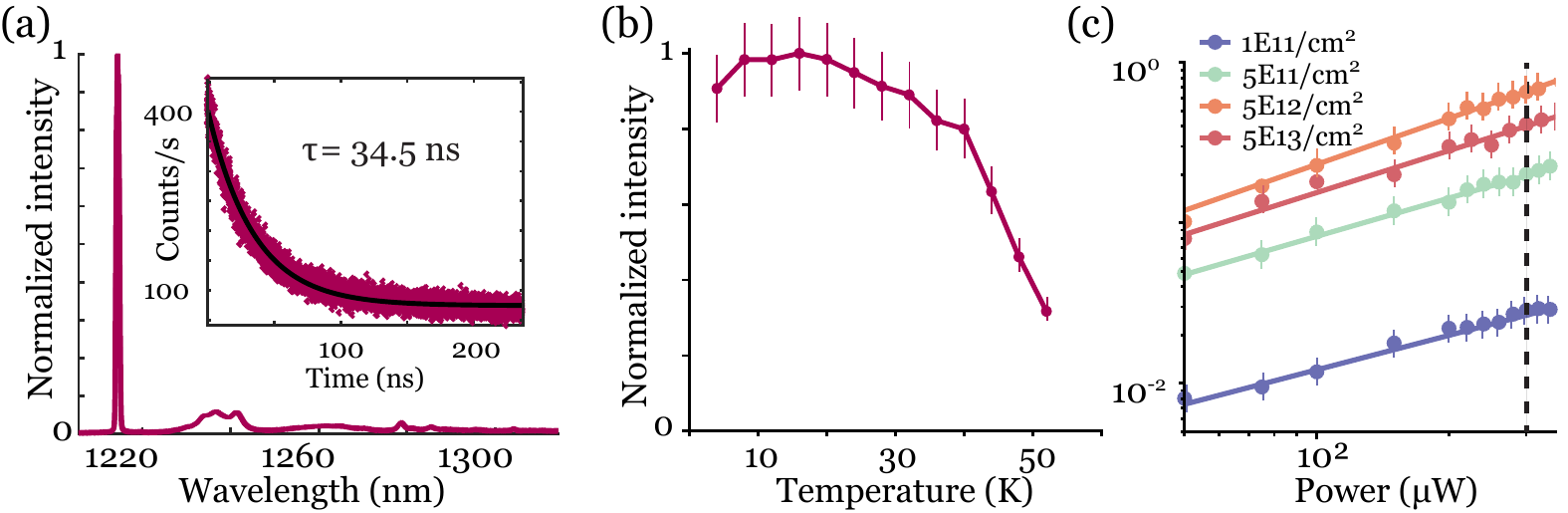}
\caption{\label{fig.characterization} (a) Spectrum and lifetime measurement for W center implanted at 80 keV in bulk $p$-type silicon (wafer type A). (b) Temperature dependence of PL intensity for W center implanted at 80 keV in bulk $p$-type silicon (wafer type A). (c) Pump power dependence for W centers implanted in SOI with different fluences (wafer type C). Straight lines are linear fits on a log-log plot. The vertical dashed line indicates pump power used in optimization studies.}
\end{figure}

Figure \ref{fig.characterization} (a) shows a typical spectrum from the $p$-type bulk silicon (wafer type A) implanted at 80 keV with a fluence of 5$\times 10^{13}$ /cm$^2$. The time-dependence of the PL when pumped with a 672 nm pulsed laser with a 1 MHz repetition rate is shown in the inset. A fit to the decay indicates a total (radiative and non-radiative) lifetime of (34.5 $\pm$ 0.5) ns (the error is estimated from the standard error of the fit). For the lifetime measurement, light from the sample was fiber-coupled to a superconducting-nanowire single-photon detector and correlated with the pump laser trigger.  This lifetime is fast enough for superconducting optoelectronic neuromorphic computing \cite{shbu2018}. For other applications that require higher speed, the lifetime may be decreased through engineering of the local density of optical states \cite{Sumikura2014} or by using other silicon-based emitters with faster intrinsic lifetime \cite{Beaufils2018, Chartrand2018}. The W center PL also exhibits a strong temperature dependence, with the PL intensity decreasing sharply around 45 K, as shown in Fig. \ref{fig.characterization} (b). The data also shows a decrease in PL intensity for the point at 5 K. This drop in PL intensity at low temperatures has been observed previously to be sample dependent \cite{Davies1989}, and can be explained by free carriers becoming captured in shallow traps at very low temperatures. For operation below 4 K, as required for SNSPDs, this is a concern that should be addressed with measurements at lower temperatures and with different substrates. Measurements presented later in this work were performed in a cryostat with a minimum temperature of around 20 K as the PL intensity is relatively flat in this region (see Appendix B).  The error bars given throughout the paper are two standard deviations (2$\sigma$), where the standard deviation is a fixed percentage of the mean calculated from repeated measurements of certain samples. If several chips from the same wafer were measured in one cool down, the standard deviation over that cool down was used to calculate the error bars. A discussion of the sources of error and the error bars throughout the paper is given in Appendix B.

A known property of emitters such as quantum dots and semiconductor defects is that the PL intensity saturates at high pump power. To compare the PL intensity of different samples, it is important to operate in the non-saturating regime. Figure \ref{fig.characterization} (c) shows the PL intensity versus pump power for different fluences implanted in SOI (wafer type C) at an energy of 25 keV. The solid lines show fits to the data. The slopes of the fits in Fig. 1 (c) (see Appendix B) are slightly less than 1 for most of the samples. The slope increases with higher fluence, suggesting that lower fluences may start to saturate by 300 \textmu W. We note that the saturation power is high since we are measuring ensembles of W centers. The saturation power is likely much lower for a single W center.

\section{Depth dependence}
\label{sec.depth}

\begin{figure}
\centering\includegraphics[width=11.4cm]{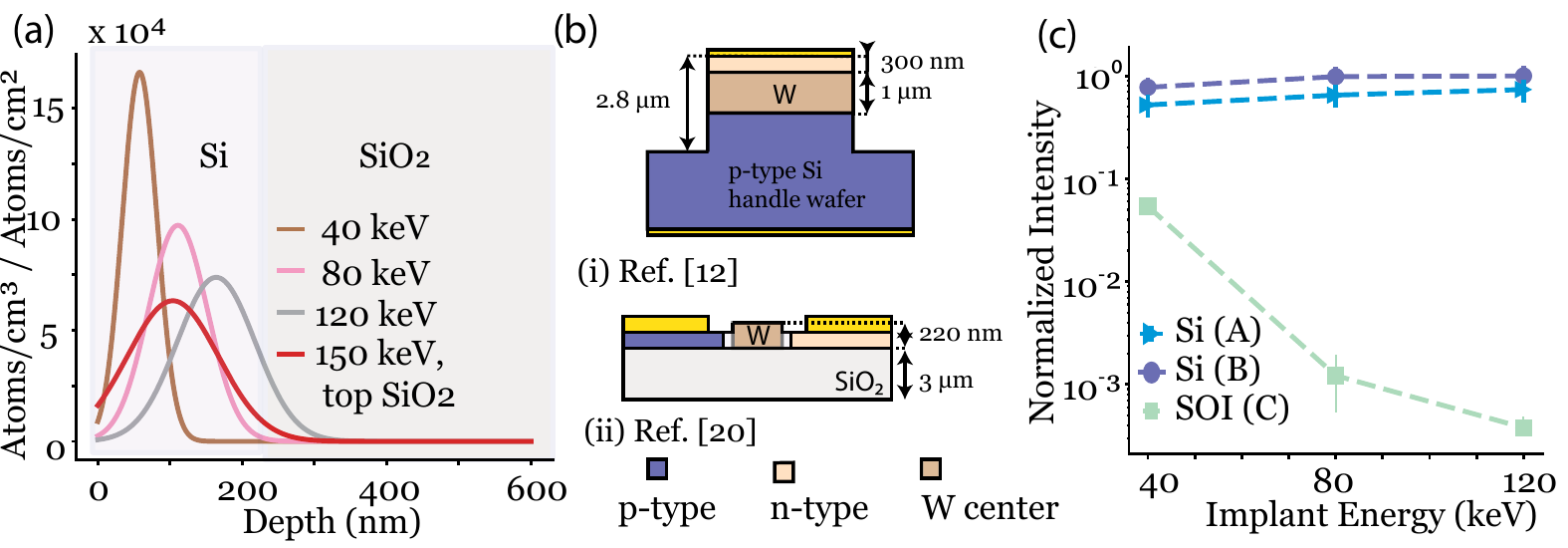}
\caption{\label{fig.depth}(a) Simulated stopping range of $^{28}$Si$^+$ ions in Si for the three different energies shown in part (c), as well as under the conditions in Ref.\,\cite{Buckley2017}. (b) Schematic of the geometry of the LEDs in Refs. \cite{Bao2007} and \cite{Buckley2017}. (c) PL from W centers at three different energies in three different wafer types (see text). }
\end{figure}

Ions implanted in a substrate will have an average projected range dependent on the energy of the implant, as shown in Fig. \ref{fig.depth} (a). Many other atoms in the silicon lattice are displaced during the implantation. The displaced and recoiled ions can settle as interstitials well beyond the projected range, leaving vacancies in their paths. The post-implant anneal allows these interstitials to migrate and form W centers. The intensity of the PL depends on a competition between the rate of the radiative recombination at the W center with the rates of other non-radiative processes at other defects \cite{Harding2006, Aboy2018, Bao2007}.  A review of the literature (see Table 1) indicates that while W center formation peaks at the projected range of implanted ions, PL intensity peaks much deeper. The literature survey also indicates that there is a strong fluence and energy dependence on the depth at which the PL intensity is maximized, and further study is needed for optimization in a particular substrate, such as SOI.

The W center is thought to be a particular configuration of $n=3$ clusters of silicon interstitials\cite{Aboy2018}. Many other clusters of interstitials and voids are formed during the implantation process. Therefore, it is not only the total number of W centers that are formed that is important, but the number of W centers relative to other centers. In particular, there is evidence that $n>5$ clusters strongly quench the W center PL \cite{Aboy2018}. The strong dependence of stopping range on implant energy and the complicated dependence of the number of clusters of $n$ silicon interstitials on implant conditions and depth suggests that it is important to investigate the PL intensity dependence on the energy and fluence of the implanted ions for a 220 nm thick device layer, the most common choice for integrated photonics using SOI.  

The stopping ranges for silicon in silicon, calculated using the software SRIM \cite{srim}, are shown in Fig. \ref{fig.depth} (a) for implant energies of 40 keV, 80 keV and 120 keV, as well as 150 keV implanted through a 100 nm top oxide. Silicon ions implanted at 80 keV have a projected range of 110 nm (the center of 220 nm SOI). 80 keV is also the energy that was used in the LED in reference \cite{Bao2007}. This LED consisted of a vertical $p$-$i$-$n$ diode with a laser annealed top surface, as shown in Fig. \ref{fig.depth} (b) part (i). W centers were implanted with a fluence of $10^{15}$ /cm$^2$. The $i$-region, where the W centers should be located for optimal performance, was 1 \textmu m wide. The $p$-region, located vertically above the $i$-region, and through which the W centers were necessarily implanted, was 300 nm wide. This means the center of the W center distribution should be $\approx$ 800 nm deep for optimal performance. Earlier studies \cite{Skolnick1981,Gotz1984,Giri2001} (Table \ref{table.depth}) indicated that under similar implantation and annealing conditions the W centers were mostly located much deeper than the projected range (110 nm), with estimates of up to 500 nm. Table \ref{table.depth} summarizes the previous work discussing depth dependence. The `Estimated W center PL depth' column in Table I refers to the depth of the center of the W center distribution estimated from etch-back studies of the PL intensity. The `desired depth' column in Table I refers to where the center of the W center distribution would be located for optimal device characteristics (e.g. the center of the $p$-$i$-$n$ junction for an LED). Figure \ref{fig.depth} also includes the stopping range profile for silicon implanted with 150 keV energy through a 100 nm capping oxide. This is the implant condition in Ref.\,\cite{Buckley2017} where waveguide-coupled LEDs were fabricated in 220 nm device layer SOI (geometry shown in Fig. \ref{fig.depth} (b) (ii)). The projected range was designed to be 110 nm, the center of the silicon device layer, when the top oxide layer is removed. However, no etch-back study of the PL was performed in this case, and based on the prior work this is likely not the optimal depth for W center PL, as will also be demonstrated later in this paper.  

Based on the mechanism discussed at the start of this section for W center luminescence \cite{Aboy2018}, the depth dependence could be explained as follows. Despite the fact that W centers are formed at the highest density near the projected range, the $n>5$ silicon interstitial clusters cause the W center PL to be quenched until much deeper. These $n>5$ clusters are also formed at the highest density near the projected range, but their numbers fall off more steeply with increasing depth. An alternate explanation involving a competitive process with voids has also been proposed \cite{Harding2006}, but the depth dependence is not clearly explained by this process. 

\begin{table}
\caption{\label{table.depth}Summary of previous work}
\resizebox{\textwidth}{!}{
 \begin{tabular}{| l | c | c | c | c | c | c | c |}
 \hline
 Ref. & Species & Energy (keV)& Fluence & Anneal & Projected range & Estimated W center PL depth & Desired depth \\ \hline
\cite{Buckley2017} & $^{28}$Si${^+}$ & 150 & 5$\times$10$^{12}$ & 250$^\circ$C, 30 minutes & 210 nm & n/a & 210 nm \\
\cite{Bao2007} & $^{28}$Si${^+}$, $^{30}$S$^{+}$ & 80 & 10$^{15}$ (Si),10$^{14}$ (S) & laser anneal 1.4 J/ cm$^2$, 275$^\circ$C 30 min & 110 nm & $>$200 nm, $<$2 um & 800 nm \\
\cite{Giri2001} &  $^{28}$Si${^+}$ & 80 & 10$^{13}$ & implanted at 265$^\circ$C & 110 nm & $<$200 nm & n/a \\
\cite{Giri2001} &  $^{28}$Si${^+}$ & 80 & 5$\times$10$^{13}$ - 10$^{15}$ & implanted at 265$^\circ$C & 110 nm & $>$ 255 nm, $<$ 1\textmu m & n/a \\
\cite{Giri2005} & $^{28}$Si${^+}$ & 80 & 10$^{15}$ & implanted at 260$^\circ$C & 110 nm & 300 nm & n/a \\
\cite{Gotz1984} & As$^{+}$ & 100 & 5$\times$10$^{15}$ & laser anneal (1.75 J /cm$^2$) & 100 nm & 300 nm & n/a \\
\cite{Skolnick1981} & $^{29}$Si$^+$ & 80 & 3$\times$10$^{15}$ & laser anneal (1.5 J /cm$^2$) & 110 nm & 500 nm & n/a \\ \hline
\end{tabular}}
\end{table}

To test this depth dependence, $^{28}$Si$^{+}$ ions were implanted at 40 keV, 80 keV and 120 keV in the three different substrates through a screen oxide of around 5 nm (5 nm to 7 nm measured). Implants were made on full wafers with a fluence of 5$\times 10^{13}$ /cm$^2$. The initial results are shown in Fig. \ref{fig.depth} (c). We observe that for both the $p$-type bulk silicon (wafer type A) and intrinsic bulk silicon (wafer type B) there is not a strong dependence on the implant energy. The intrinsic silicon (wafer type B) is brighter by a factor of around 1.5 than the low resistivity $p$-type silicon (wafer type A). It has been noted previously that boron doping has a quenching effect on the PL \cite{Charnvanichborikarn2010}, and this is likely the cause of the difference in PL intensity. However, there is a strong energy dependence in the SOI sample. We observe that SOI C has a maximum PL intensity for the lowest (40 keV) energy. 

\section{Optimization}
\label{sec.optimization}

\begin{figure}
\centering\includegraphics[width=11.4cm]{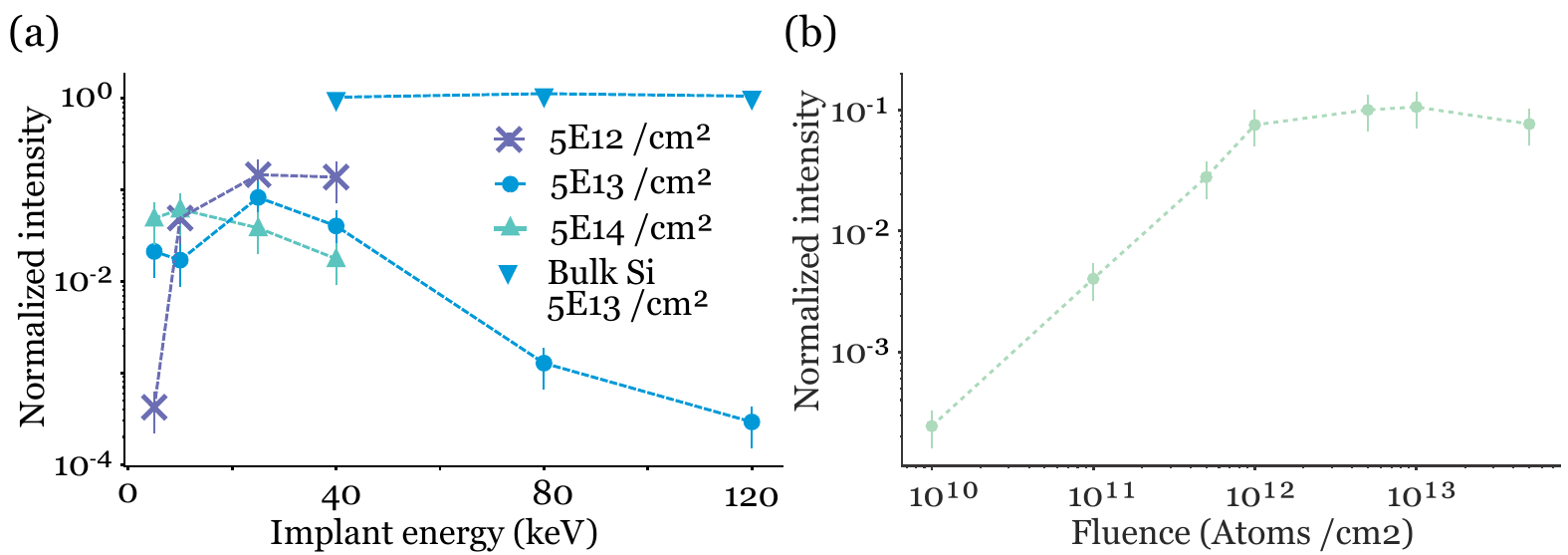}
\caption{\label{fig.optimize-energy-fluence} (a) PL intensity versus implant energy for SOI C. (b) PL intensity versus fluence for SOI C for an implant energy of 25 keV.}
\end{figure}

The data shown in Fig. \ref{fig.depth} and Table I indicates the potential for further optimization of the energy, fluence and anneal conditions for the implants, particularly for SOI wafers where the silicon layer thickness is chosen to be 220 nm for waveguide formation. For this study, a single SOI wafer (wafer type C) was diced into 1 cm die, which were attached to silicon carrier wafers with poly(methyl methacrylate). Implants were then performed on these die. After implantation, each 1 cm die was removed from the silicon carrier wafer, and diced into 2.5 mm die so the PL from multiple die could be compared in a single cooldown. The 2.5 mm die were annealed in batches for a single experiment. The remainder of the measurements were performed in a custom-built continuous-flow cryostat with optical access. This cryostat operated between 24 K and 32 K.  As can be seen from the temperature dependence plot in Fig. \ref{fig.characterization} (b), this is close to the optimum temperature for PL intensity from W centers. The optical setup for the remainder of the experiments used a 0.42 NA 50$\times$ objective. The measurements are normalized to the bulk $p$-type silicon sample implanted at an energy of 40 keV and fluence of 5$\times 10^{13}$ /cm$^2$. 

Figure \ref{fig.optimize-energy-fluence} (a) shows further optimization of the energy of implantation for three different fluences (5$\times$10$^{12}$ /cm$^2$, 5$\times$10$^{13}$ /cm$^2$, 5$\times$10$^{14}$ /cm$^2$). We observe the optimum conditions in PL intensity for an implant energy of 25 keV and fluence of 5$\times 10^{12}$ /cm$^2$. The intensity is still only around one tenth of the PL intensity of the 40 keV, 5$\times 10^{13}$ /cm$^2$ bulk silicon. We also observe that the optimum energy appears to change with fluence for the three fluences measured. To further bracket the optimal fluence, we measure the PL intensity for eight different fluences implanted at an energy of 25 keV. The results for SOI C are shown in Fig. \ref{fig.optimize-energy-fluence} (b). The PL intensity increases with increasing fluence from 10$^{10}$ /cm$^2$ to 10$^{12}$ /cm$^2$, before saturating and then decreasing at higher fluences. Fitting the linear part of this curve we find a slope of 1.2$\pm$0.02 (where the error is the standard error in the slope) on a log-log plot, indicating that the PL intensity is nearly proportional to the fluence. The slope is likely larger than 1 due to the fact that the PL from lower fluences are not exactly proportional to the pump power (starting to saturate), as discussed in Appendix B. The exact fluence at which the PL intensity peaks depends on the energy of the implant, indicated in Fig. \ref{fig.optimize-energy-fluence} (a). This saturation of intensity with fluence has also been previously observed in the literature \cite{Harding2006} but at much higher implant energies of 1 MeV. That study found that the W center PL was only proportional to fluence for fluences from $10^8$ /cm$^2$ to $10^{10}$ /cm$^2$, two orders of magnitude lower than in this study. This finding is consistent with the trend in Fig. \ref{fig.optimize-energy-fluence} (a), where it appears that the optimal fluence is lower for higher energy implants. This is likely due to the fact that the number of $n$-interstitial clusters formed is a non-linear function of fluence, with a larger ratio of high-$n$ to low-$n$ clusters formed at high fluences (for the same implant energy). This leads to the optimal ratio for PL occuring at a larger depth \cite{Aboy2018}.

\begin{figure}
\centering\includegraphics[width=11.4cm]{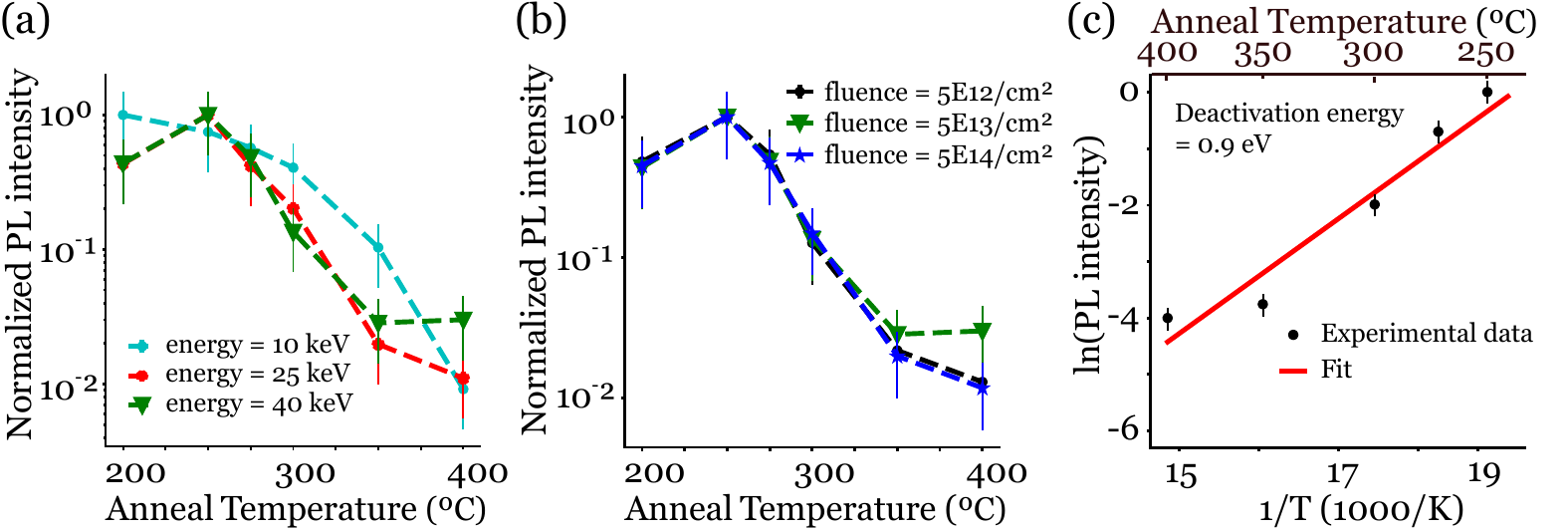}
\caption{\label{fig.anneal} (a) PL intensity versus anneal temperature for SOI C for three different energies with a fluence of 5$\times 10^{13}$ /cm$^2$. (b) PL intensity versus anneal temperature for SOI C implanted at 40 keV for three different fluences. Data in parts (a) and (b) are normalized relative to the max value at those implant conditions. The dashed lines are guides to the eye. (c) Arrhenius plot of the data from part (b).}%
\end{figure}

Next we consider annealing conditions. It has been reported \cite{Giri2001} that an anneal temperature of 265$^\circ$C gives the optimal PL intensity. However, the test in Ref.\,\cite{Giri2001} was performed in bulk silicon for an implant energy of 80 keV and a fluence of 5$\times 10^{13}$ /cm$^2$. Due to the competitive nature of the PL process, we considered that the anneal conditions might also depend on the fluence or energy of implantation. Therefore we annealed different samples at different temperatures to see if the peak in PL intensity versus anneal temperature  depends on implantation fluence or energy shifts for different samples. This is shown in Fig. \ref{fig.anneal} (a) and (b) for different energies and fluences. Unlike Figs. 2 and 3, in this case the data has been normalized to the maximum intensity for that implant condition. The relative intensities of the peaks can be seen in the previous figure (Fig. \ref{fig.optimize-energy-fluence}). The peaks do not appear to be significantly different, suggesting that there is no strong dependence on fluence or energy for the optimal anneal conditions.

The annealing data can be used to extract the activation and deactivation energy of the W center formation process. The activation and deactivation energy refer to the energy of the rate limiting step in the formation and decay of the photoluminescent W center. The deactivation energy is most likely the decay energy of the W center cluster, but it could alternatively be the energy of formation of a strong competitive non-radiative center. Reference \cite{Schultz1992} found that the activation energy of the W center is $0.95\pm0.05$ eV and the deactivation energy is $1.2\pm0.05$ eV for W centers implanted at a fluence 4$\times$10$^{12}$ and energy 1 MeV \cite{Schultz1992} . A similar study \cite{Giri2001} found an activation energy of $0.85\pm0.05$ eV for an implant energy of 80 keV and fluence of 5$\times$10$^{12}$ /cm$^2$, which also indicates that the activation energy is stable over a wide range of implant energies (deactivation energy unreported). The fact that previous studies have found the activation and deactivation energies to be very similar to each other suggests that it is the energy for the formation and decay of the center itself. A thorough discussion of the activation and deactivation energies of the W center is found in Ref.\,\cite{Schultz1992}, although the precise mechanism for the formation/decay process is still unknown. The (de)activation energy is found from the slope of the fit to an Arrhenius plot\cite{Pauling1988} of $ln{(k)}$ versus $1/T$. The equation of the fit is $\ln{(k)} = \frac{E_D}{R}(\frac{1}{T})+\ln(A)$, where $E_D$ is the deactivation energy, $T$ is the anneal temperature, $R$ is the gas constant, and $k$ is the PL intensity. $A$ is a constant for the W center formation process. We show this fit in Fig. \ref{fig.anneal} (c) for the deactivation energy, where we obtain an average deactivation energy of (0.9$\pm$0.1) eV. The error is calculated from the standard error in the slope. The errorbars do not quite explain the deviation of the data from the fit. A detailed discussion of the calculation of the errorbars is given in Appendix A and B. The data points in Fig. \ref{fig.anneal} (c) are averages over the data points in Fig. \ref{fig.anneal} (b), but we have also calculated the activation energies from the slopes of the curves for the individual fluence and energy measurements in Figs. \ref{fig.anneal} (a) and (b). The minimum deactivation energy measured was (0.8$\pm0.1$) eV for an implant energy of 40 keV and a fluence of 5$\times10^{13}$ /cm$^2$. The maximum calculated deactivation energy was (1.0$\pm0.1$) eV for an implant energy of 25 keV and fluence of 5$\times10^{13}$ /cm$^2$. This variation is likely measurement error, as there was no trend with either energy or fluence in the calculated value. We do not have sufficient data for low anneal temperatures to fit the activation energy. We note that the 10 keV sample shows an anomalous anneal curve shape, but it is not clear if this is significant. 

To produce electrically injected optical devices, it is necessary to mask off the implants that produce W centers. We therefore studied the masking properties of several resists. Three additional silicon wafers were implanted at 25 keV and fluence 5$\times$10$^{12}$ /cm$^2$. The wafers had 600 nm thick electron-beam resist, 1 $\mathrm{\mu}$m thick photoresist and 3 $\mathrm{\mu}$m thick photoresist. We found that the electron-beam resist was fully removed with 30 minutes heated N-methyl-2-pyrrolidone followed by 10 minutes soak in a sulfuric acid solution. The photoresists were fully removed with 5 minutes sonication in acetone followed by 2 minutes sonication in isopropyl alcohol and 10 minutes soak in a sulfuric acid and hydrogen peroxide solution. We did not observe PL from the areas of silicon that were masked by any of the resists, which allows us to use any of these as a lithographic masks for the implants. Finally, it was reported \cite{Schultz1991} that high temperature implants in silicon lead to significantly less lattice damage and consequently fewer nonradiative channels. To test how this affects PL from the W center, a final wafer was implanted at 265$^\circ$C (and not subjected to a post-implant anneal), with an energy of 40 keV and fluence 5$\times10^{13}$ /cm$^{2}$. The hot implanted wafer showed a factor of two increased PL over a wafer annealed post-implant. While this is promising for future improvements, more advanced fabrication is necessary to mask hot implants, as typical photoresists cannot withstand processing at this temperature.

\section{W center light sources with 300 mm CMOS-friendly processes}
\label{sec.foundry}

\begin{figure}
\centering\includegraphics[width=11.4cm]{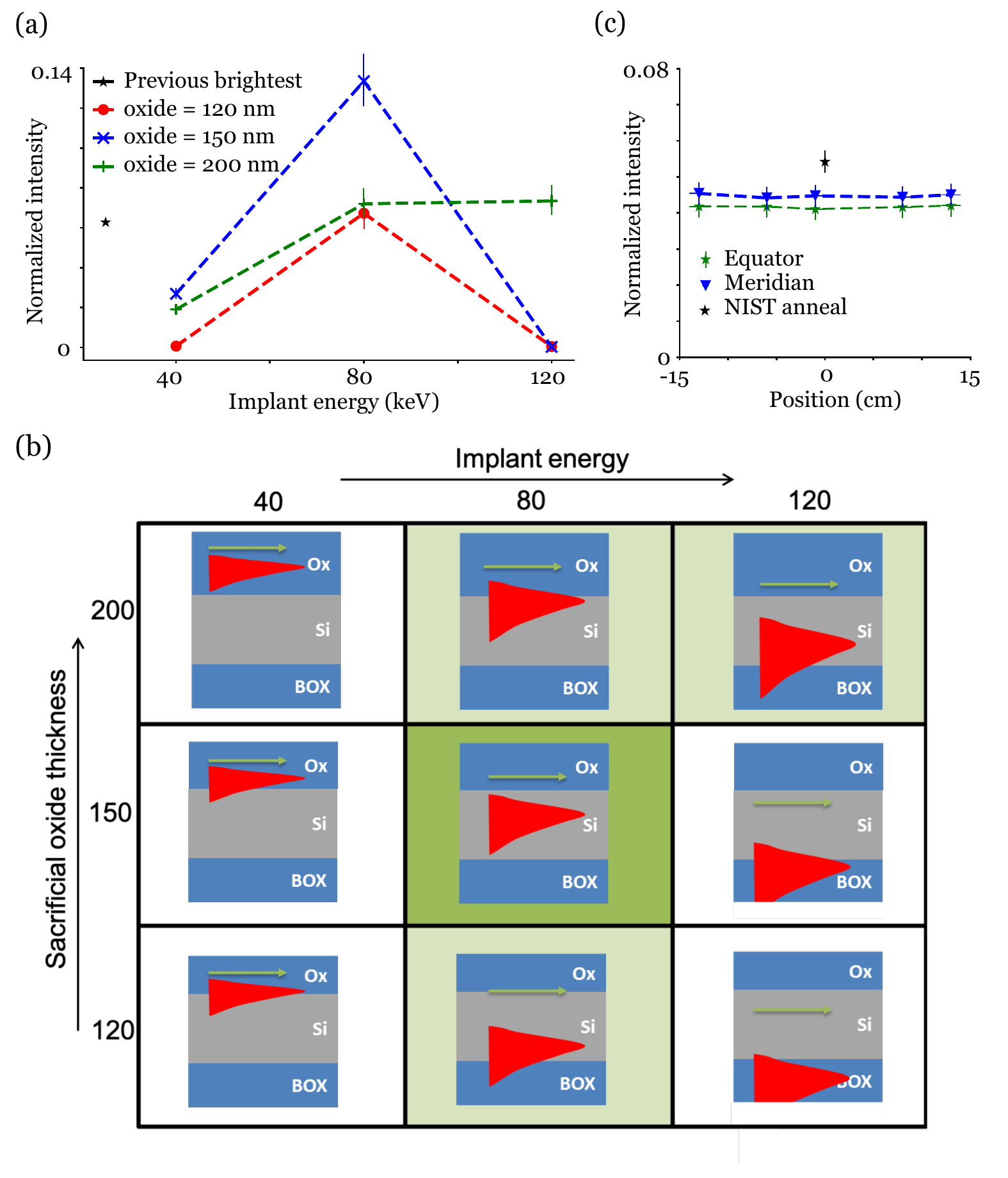}
\caption{\label{fig.suny} (a) PL intensity versus energy for 300 mm SOI wafers implanted through different barrier oxide thicknesses. (b) Schematic showing the oxide-silicon-oxide structure being implanted in part (a), with the blue representing silicon dioxide and the gray representing silicon. The green arrow shows the range of silicon ions in this structure, while the red shows the presumed depth where the W center PL is originating based on the results of part (a). (c) PL intensity versus position along the equator or meridian of a 300 mm wafer processed entirely at SUNY. The NIST annealed sample is from a wafer implanted under the same conditions but with the anneal performed at NIST.}%
\end{figure}

To generate low-cost on-chip light sources, these devices must be fabricated in a conventional foundry process. We take the first steps towards this goal by demonstrating that the implantation can be done at a 300-mm-wafer scale. These implants were done at the cleanroom at the State University of New York (SUNY) Polytechnic Institute. In this study, ions were implanted at 40 keV, 80 keV and 120 keV at an angle of zero degrees to normal, with a fluence of 5$\times10^{13}$ /cm$^{2}$, in 300 mm SOI wafers with buried oxide thickness of 145 nm. The implants were performed through a thick barrier oxide, with the intention of leaving the high-damage sections in the oxide and removing them with hydrofluoric acid after annealing. The oxide deposition, anneal and oxide removal were all performed on full wafers using standard 300-mm process tools at SUNY Poly. Ten wafers were implanted at three different implant energies (40 keV, 80 keV and 120 keV) through three different oxide thicknesses (120 nm, 150 nm and 200 nm). The fluence was again 5$\times$10$^{13}$ /cm$^2$. In addition, a single wafer was implanted with a 5 nm screening oxide at 40 keV, still at zero degree tilt. PL measurements were performed in the same method as in the previous section, and normalization was done relative to the same bulk silicon sample. 
Because the top layer of the wafer contains significant damage after implantation, we hypothesized that a thick barrier oxide could capture that high damage region for later removal. PL from high-fluence-generated W centers has been observed to increase in brightness after etching away of the top layer \cite{Giri2001,Schultz1992,Skolnick1981}. Figure \ref{fig.suny} (a) shows the results of this study. We observed the peak PL intensity from the wafer implanted at 80 keV through a 150 nm barrier oxide. We also cooled down the previous brightest sample, (25 keV, 5$\times 10^{12}$ /cm$^2$ in SOI C) in the same cooldown and found an increase in brightness of a factor of two relative to the previous best implant condition. The underlying substrate begins 515 nm below the top layers, and to ensure that the sample W centers were not in the handle wafer, the device layer was etched away and the samples remeasured. We did not observe any PL from the samples after removal of the device layer.

Fig. \ref{fig.suny} (b) is a schematic describing the results in Fig. \ref{fig.suny} (a). All sizes are meant as representations and are not to scale.  The schematic is split into a grid showing a subfigure for each screening oxide and implant energy. The blue regions represent silicon dioxide while the grey region represents silicon. The green arrow shows the projected range of the silicon ions while the red shows our estimates of where the PL is predominantly originating. For low energies and thick oxides, the oxide prevents the formation of W centers (i.e. the red region of the preferred W center depth is almost entirely in oxide). For high energies and thin oxides, the W centers are likewise cut off by the bottom oxide. For higher energies we also expect the distribution of W centers to be broader than for lower energies, which is represented by the width of the red W center region.

A primary goal of this portion of the study was to test the uniformity of W center creation across the 300 mm wafer. The wafer implanted at 40 keV through a 5 nm barrier oxide was diced into 1 cm squares, and die were selected from equidistant points along the equator and meridian. Each of these 1 cm die was then diced into 2.5 mm die. The measurement was performed with three 2.5 mm die from each 1 cm square. The variation in the measurements from within 1 cm was compared to the variation across the entire wafer, with a total of 15 2.5 mm die compared in each cooldown (Fig. \ref{fig.suny} (b)) No significant variation was observed across the wafer (equator standard deviation 4\%, meridian standard deviation 11\%). Interestingly, die from a previous wafer implanted with the same conditions but annealed at NIST showed consistently higher PL intensity (labeled NIST anneal on Fig. \ref{fig.suny} (c)). This discrepancy may be explained by a difference in furnace temperature calibration. We observe approximately 4\% variation in PL per degree around the optimal anneal temperature. Therefore this discrepancy could be explained if the anneal chambers were inadvertently operating at slightly different temperatures. The difference between the equator and meridian samples is due to the fact that the error between cooldowns is higher than the error between samples measured in the same cooldown. This is possibly due to differences in the temperature of the cryostat or the optical apparatus.

\section{Further work}
\label{sec.future}

We have investigated the PL intensity from W centers implanted in the standard SOI substrate used for silicon integrated photonics, with a 220 nm device layer thickness. We find that the PL is strongly dependent on the energy of the implant. We find optimal photoluminescence for the implant conditions of 5$\times$10$^{12}$ /cm$^2$ fluence and 25 keV energy. 

Based on the data presented here, it is unlikely that previous attempts to use the W center as a light source have used optimal implant and anneal conditions. It may be possible, using some of the results of this study, to significantly improve the brightness of W center based silicon light sources. While there have been demonstrations of cavity-coupled PL from silicon defects in the past \cite{LoSavio2011, Sumikura2014}, a direct comparison of the brightness of these emitters to the W center has not been made. It is possible that the W center could demonstrate much stronger cavity-coupled luminescence. The one previous known attempt to cavity couple the W center luminescence used suboptimal implant conditions (implanting with 100 keV energy for a membrane thickness of 220 nm) \cite{Kuznetsova2012}. For applications in superconducting optoelectronic neuromorphic computing \cite{Shainline2017a,shbu2018}, total light-production efficiency of 10$^{-4}$ for LEDs operating at 4K is sufficient to enable power-efficient, large-scale systems.  In that context, light production efficiency greater than 1\% provides little advantage \cite{shbu2019}, as photon detection with superconducting detectors dominates the energy budget at that point. The previous demonstration with all-silicon waveguide-integrated LEDs showed a system efficiency of 5$\times$10$^{-7}$. If the optimal conditions remain the same in the case of electroluminescence, we expect an efficiency of 5$\times$10$^{-5}$ by only changing the implant conditions. Significant further gains may be achieved through a combination of improved electrical injection and improved coupling of the W centers to the optical mode.

It remains to be seen how the implant conditions affect electroluminescence in LEDs. It has been observed that the series resistance in LEDs increases with fluence in the LEDs. Therefore, it is possible that there is a tradeoff in electroluminescence intensity for higher fluence. It is also possible that if the W centers are not uniformly distributed in depth in the device layer, current will preferentially flow through regions with a lower density of W centers due to decreased resistance, leading to a trade-off between coupling to the optical mode and electrical injection that is not present in the PL case.

Beyond W centers, there are numerous other luminescent centers\cite{Davies1989} in silicon, and some of these may be brighter or more suitable in other ways for various applications. A systematic study of the relative brightness of these centers in silicon has not yet been performed. It is also likely that the implant conditions for these centers must also be optimized. For example, electroluminescence of the G-center in silicon has been observed \cite{Cloutier2005}, and G-centers have been fabricated via ion implantation of C followed by proton irradiation \cite{Berhanuddin2012}. Meanwhile, studies on the depth dependence of the G center PL \cite{Skolnick1981} have indicated that the G centers are formed at even larger depths relative to the projected range than the W center. This indicates that for fabrication of G center LEDs in SOI via ion implantation, a similar study to this paper is required.

There has also been renewed interest in defects in silicon as solid-state spin qubits. While solid-state spin qubits are highly stable with extremely long coherence time, coupling of these qubits has remained a challenge. Photonically addressable spin qubits can provide a method for scaling to quantum networks \cite{Ortu2018}.  Isotopically pure silicon is a strong candidate for these networks, if a photonically addressable spin qubit can be found. There has been recent interest in chalcogenide  \cite{Morse2017} and magnesium \cite{Abraham2018} defects coupled to cavities for these cavity-QED applications. The spin properties of the W center have not been fully examined \cite{Davies1987}, and it remains to be seen if W centers can act as single photon sources. An on-chip electrically injected single photon source that could be easily coupled to silicon photonic integrated circuits would also have a variety of applications in quantum optics.

\appendix
\section{Measurement methods}

Initial measurements were made in a commercially purchased 4K closed cycle helium cryostat (cooldowns -9 to -1). We label cool downs made in this cryostat with a negative number. Later (cooldown 1 onwards), a second cryostat was built that cools down to a minimum temperature of 20K. The measurement technique in this cryostat was standardized as follows. Up to eighteen 2.5 mm die were glued to a sample mount using rubber cement. A bulk control sample 5 mm $\times$ 5 mm, 40 keV from wafer A (Si $p$-type 1-10 $\Omega$ cm bulk) was also glued to the sample mount. This was the same sample every time until cooldown 20, at which point the cryostat died and was replaced, and a different sample from the same wafer was used. A Si detector was used to measure the input pump power at a point before the objective. The power was set to (300 $\pm$ 10) \textmugreek W. A spectrum was taken of the bulk sample on an InGaAs spectrometer array, and the PL intensity was optimized. The intensity was then recorded with integration time of 0.5 s averaged 5 times. The samples of interest were then measured in the same way, but with an integration time of 2 s. A second measurement of the bulk sample was then taken before warming up. The reported PL intensity is the peak count number in the zero phonon line per second, divided by the mean value of the control sample per second. The following subsection describes the variability in our results.

\section{Consistency of results}
\label{sec.Consistency of results}

\subsection{Summary of cooldowns and reproducibility}

\begin{figure}
\centering\includegraphics[width=11.4cm]{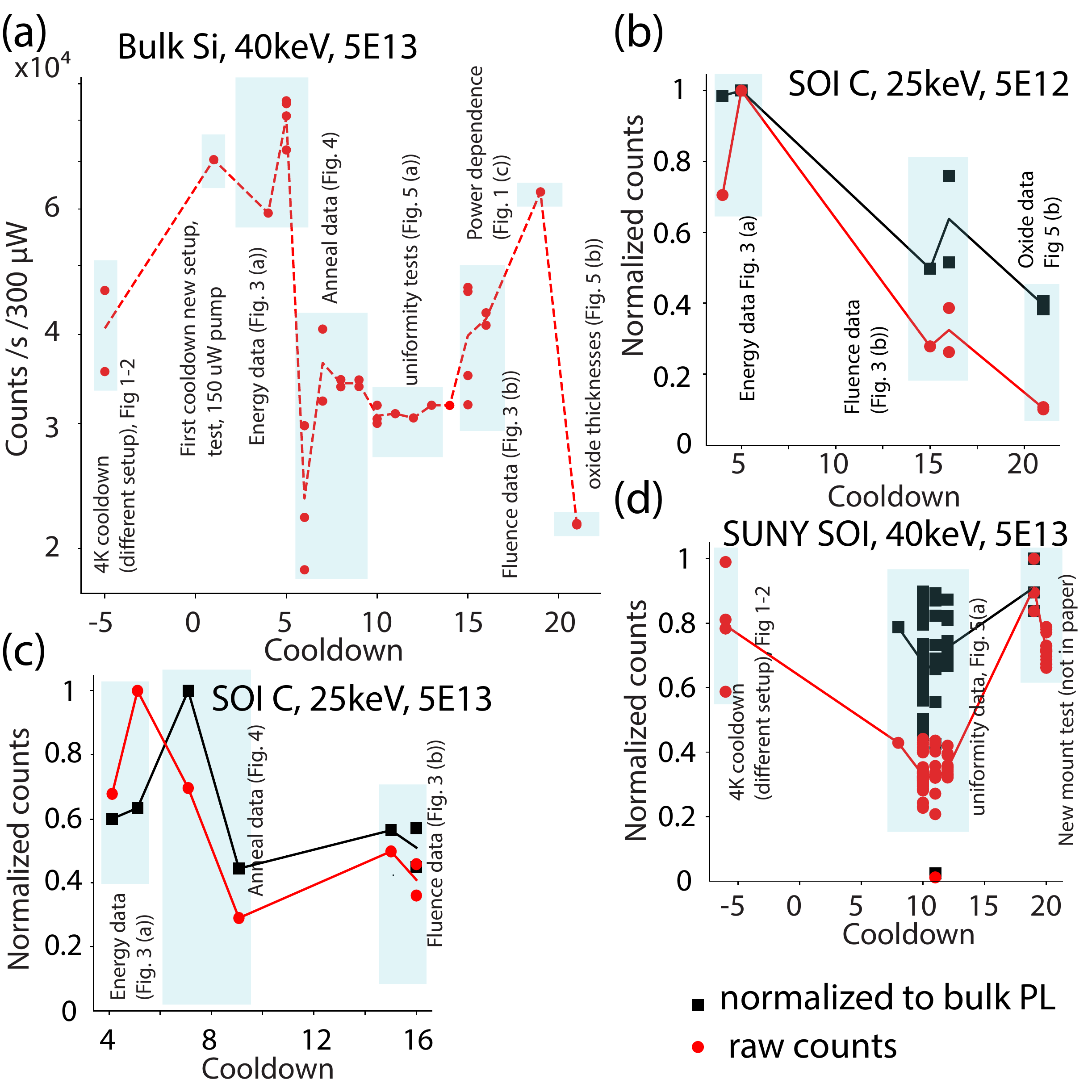}
\caption{\label{fig.uniformity} (a) Measurements of the 40 keV 5E13$/$cm$^2$ p-type bulk Si control sample taken across 26 recorded cool downs between October 2018 and August 2019. (b) Repeated measurements of SOI C 25keV  5E12$/$cm$^2$ samples. The samples in cool down 5 were implanted in a different wafer run than those in the subsequent cool downs. (c) Repeated measurements of SOI C 25keV  5E13$/$cm$^2$ samples. (d) SUNY Poly SOI sample 40keV 5E13$/$cm$^2$.}
\end{figure}

Table 2 summarizes the cooldowns that comprised the results in this paper. Noteworthy events are the switching of the measurement cryostat (transition from negative to positive cool down number) and the replacement of the cryocooler at cool down 20. Figure \ref{fig.uniformity} (a) shows the variability in the measurements of the control sample. The dots indicate individual measurements while the dashed line tracks the mean. The mean value of the PL intensity was 41000 counts/ s/ 300 \textmugreek W, the median was 34000 counts/ s/ 300 \textmugreek W and the standard deviation was 18500 counts/ s/ 300 \textmugreek W, or 50\% of the mean. The maximum deviation from the mean was $106\%$. The maximum deviation from the mean (of that cool down) in a single cool down was $27\%$ (max standard deviation 20\%). We would therefore expect that if we normalize to the mean of the control counts on a particular cooldown, the answer may vary by at most $\approx$ 50\%. If we don't normalize, we would expect the result to vary by a similar amount to the control, so up to $\approx$ 100\% variation.

There were also several implant conditions that were measured multiple times. Three of these are shown in parts (b)-(d).  In each of these cases the values were normalized to the maximum measured values (red) and the mean value of the bulk PL sample measured in that cool down (measured value/mean bulk PL measurement). In the main paper text, this value is left as the measured value/mean bulk PL measurement, to show how close we are getting to the bulk PL, but in this case for comparison reasons the value is normalized. The standard deviations for the data shown in parts (b) to (d) are 38 (76)\%, 28 (40)\% and 20 (46)\% for the values normalized to bulk (raw counts), and matches what we expect based on the above analysis of the variation in control counts. This is the expected variation when comparing values between plots. 

However, if we limit ourselves to samples annealed together and cooled down together in one cool down, the standard deviation decreases significantly. For example, in the data in Fig.\ref{fig.suny} (c), all of the equator chips from a single cooldown had a standard deviation of 4\%, from measurements of twelve separate chips, while we obtain a standard deviation of 11\% for the meridian chips. However, if we include a wafer implanted under identical conditions but annealed at NIST, rather than at SUNY Poly, the standard deviation increases to $15\%$. This is still lower than if we include all data from this implant condition over all cool downs, when the standard deviation increases to 20\% (when normalized to a control sample, 46\% un-normalized).

Error bars on any of the plots in the main text refer to errors between measurements in that plot, and we leave that at 20\%, unless the plot contains data from multiple cooldowns. If we made multiple measurements of the same samples within the cool down and the error was higher than 20\%, we have set the error bars to this higher value. The errors between plots are expected to be higher, with up to a factor of two difference observed from measurements of the same sample on different plots. The next section describes the sources of error that may be contributing to this.

\begin{table}
\caption{\label{Cooldown table}Summary of cooldowns}
\resizebox{\textwidth}{!}{
\begin{tabular}{| l || l | l |}
\hline
 CD	& DATE & NOTES \\ \hline \hline
-5 &	11/26/2018 &	Fig. 2 (a) \\ \hline
-4 &	11/27/2018 &	Fig. 2 (a), Fig. 1 (c) \\ \hline
-3 &	11/29/2018 &	SOI C measurements versus energy \\ \hline
-2 &	12/1/2018 &	Etched SUNY sample measurements (not in paper due to on-axis implant) \\ \hline
-1 &	12/4/2018 &	Intrinsic silicon measurements. Switch to 20K cryostat \\ \hline
1 &	2/25/2019 &	First measurements in new 20K cryostat. Test/bulk silicon. 27K \\ \hline
2 &		 & No measurements \\ \hline
3 &	2/26/2019	 & Repeat/reproduce measurements Si and SOI C (Fig. 2) \\ \hline
4 &	2/28/2019 &	Fig. 3 (a) energy measurements \\ \hline
5 &	3/07/2019 &	Fig. 3 (a) energy measurements \\ \hline
6 &	3/27/2019 &	Fig. 4 Anneal measurements \\ \hline
7 &	3/28/2019 &	Fig. 4 Anneal measurements \\ \hline
8 &	3/29/2019 &	Fig. 4 Anneal measurements \\ \hline
9 &	4/4/2019 &	Fig. 4 Anneal measurements \\ \hline
10 &	4/11/2019 &	Uniformity data equator (Fig. 5 (a)) \\ \hline
11 &	4/15/2019 &	Uniformity data meridian (Fig. 5 (a)) \\ \hline
12 &	4/16/2019 &	Uniformity data meridian grease test (Fig. 5 (a)) \\ \hline
13 &	4/17/2019 &	First oxide thickness data (Fig. 5 (b)) \\ \hline
14 &	4/18/2019 &	Checking older samples \\ \hline
15 &	5/09/2019 &	Fluence (Fig. 3 (b)) \\ \hline
16 &	5/11/2019 &	Fluence (Fig. 3 (b)) \\ \hline
17 &	5/22/2019 &	Power versus fluence (Fig. 1 (c)) \\ \hline
18 &	?	 & Working on another imaging setup \\ \hline
19 &	6/07/2019 &	Test degradation with heating to 180$^\circ$C \\ \hline
20 &	6/12/2019 &	Test new chip mount. Cryostat died and replaced coldhead. \\ \hline
21 &	8/06/2019 &	Oxide thickness data (Fig. 5 (b)) \\ \hline
\end{tabular}
}
\end{table}

\subsection{Sources of errors}

\begin{figure}
\centering\includegraphics[width=11.4cm]{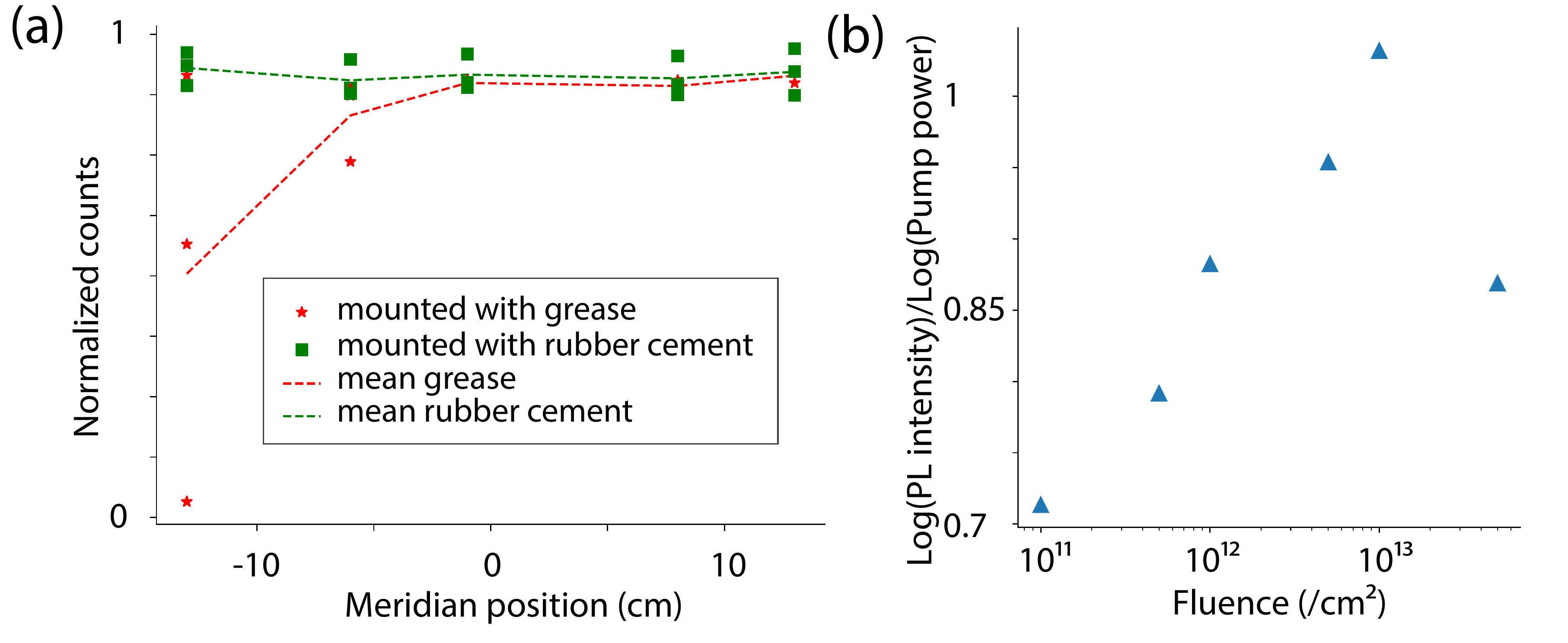}
\caption{\label{fig.errorsources} (a) The same set of samples mounted on the cold finger with rubber cement or grease. Points at the same position along the equator are taken from within the same 1 cm die at that position. (b) The fitted slope of the plot of $ \mathrm{log}$(PL intensity) versus $\mathrm{log}$(Pump power) (shown in Fig. 1 (c) in the main text) for different fluence samples implanted at 25 keV. A slope of one would indicate no saturation.}%
\end{figure}

The previous section indicates that variations in the setup can account for some of the error. In the case of a cool down, based on Fig. 1 (b) in the text there is approximately 0.5\% difference in PL intensity per degree K difference in the sample temperature region, which could rapidly increase if the temperature of the sample nears 40K. Furthermore, the heat sinking of the sample is dependent on the quality of the bond to the cold finger, as the samples are subjected to heating from the laser and blackbody radiation through the cryostat window. The importance of this bond is shown in Fig. \ref{fig.errorsources} (a) where the exact same set of samples is cooled down twice, once using rubber cement to bond the samples and the second time using vacuum compatible grease. In the case of the grease, there is clearly a much larger spread in PL intensities than in the case of the rubber cement. The standard deviation in the case of the rubber cement is 4\% while it is 28\% in the case of the grease. We attribute this to inadequate heat sinking, leading to an elevated sample temperature in some samples but not others. The cryostat itself also varies in temperature between 26 K and 32 K, which would account in a change in 4\%.

A second source of error may be related to different anneal conditions. As can be seen from Fig. 4 in the main text, there is a strong dependence of PL intensity on anneal temperature, with an approximate dependence of 4\%/$^\circ$C in the range around 250$^\circ$C. This can explain why two wafers that were nominally implanted under identical conditions can give slightly different results.

Another source of error is due to the saturation of the photoluminescence of the W centers with pump power.  In the case where there is no saturation, the PL intensity, $P_{PL}$ is proportional to the pump power, $P_{pump}$. The slope of $\mathrm{log}(P_{PL})$ vs $\mathrm{log}(P_{pump})$ (plots shown in Fig. 1 (c) in the main text) is 1. For a small amount of saturation, the slope will be less than 1. Figure \ref{fig.errorsources} (b) shows the fitted slope of this line (raw data in the main text in Fig. 1 (c)) versus fluence. We find that the slope is sublinear for low fluences, increasing for higher fluences. This suggests that for the lowest fluences, the 300 \textmu W pump power is beginning to saturate the emitters. Unfortunately we were unable to probe higher powers to see strong saturation due to power limitations of our pump laser. 

\subsection{Possibility of sample degredation}

In this study, the same sample was subjected to over 20 cool downs, and there was no statistical evidence of degradation of the sample. The variation in the PL measurement means that it cannot be ruled out, but the samples survive and produce PL for periods of several years and $>$20 cycles between cryogenic and room temperature. We also annealed three samples for 0 hours, 6 hours and 24 hours at 150$^\circ$C and saw no evidence of degradation in PL intensity.

\section*{Acknowledgments}
We thank Dr. Maria Aboy Cebrian at the University of Valladolid and Dr. Jeff Chiles and Dr. Matt Brubaker at NIST for helpful conversations and insights. We thank Mr. Ronald Bourque and his colleagues at TEL Technology Center America for help with processing at certain steps. The U.S. Government is authorized to reproduce and distribute reprints for Government purposes notwithstanding any copyright notation thereon. The views and conclusions contained herein are those of the authors and should not be interpreted as necessarily representing the official policies or endorsements, either expressed or implied, of the Air Force Research Laboratory or the U.S. Government. This is a contribution of National Institute of Standards and Technology (NIST), an agency of the U.S. government, not subject to copyright.

\section*{Funding}
SUNY Poly co-authors require the statement that this material is based on research sponsored by the Air Force Research Laboratory under agreement number FA8750-1-1-0031

\section*{Disclosures}
The authors declare no conflicts of interest.

% \begin{table}
% \end{tabular}
% \end{table}

%\begin{acknowledgments}
% Put your acknowledgments here.
%\end{acknowledgments}


\bibliography{Wcenters}
\end{document}